\documentclass[sigconf,table]{acmart}

\usepackage{booktabs} 
\usepackage{tabularx} 
\usepackage[english]{babel}
\usepackage{mathrsfs}

\usepackage{multirow}
\usepackage{multicol}
\usepackage{array}
\usepackage{arydshln}

\usepackage{amsmath}
\usepackage{amssymb}
\usepackage{booktabs}
\usepackage[inline]{enumitem}
\usepackage{siunitx}
\usepackage{stackengine}
\usepackage{extarrows}

\usepackage{caption}
\usepackage{subcaption}
\usepackage{tabularx}

\usepackage{footnote}
\makesavenoteenv{tabular}
\makesavenoteenv{table}

\usepackage{xspace} 

\newcolumntype{L}[1]{>{\raggedright\arraybackslash}p{#1}}
\newcolumntype{C}[1]{>{\centering\arraybackslash}p{#1}}
\newcolumntype{R}[1]{>{\raggedleft\arraybackslash}p{#1}}

\copyrightyear{2020}
\acmYear{2020}
\setcopyright{acmcopyright}
\acmConference[Microsoft]{}{July 2020}{Technical Report}

\author{Hamed Zamani}
\affiliation{%
  \institution{Microsoft}
}
\email{hazamani@microsoft.com}

\author{Gord Lueck}
\affiliation{%
  \institution{Microsoft}
}
\email{gordonl@microsoft.com}

\author{Everest Chen}
\thanks{This work was done while Everest Chen was affiliated with Microsoft.}
\affiliation{%
  \institution{Facebook Inc.}
}
\email{everestch@fb.com}

\author{Rodolfo Quispe}
\affiliation{%
  \institution{Microsoft}
}
\email{edquispe@microsoft.com}

\author{Flint Luu}
\affiliation{%
  \institution{Microsoft}
}
\email{flintl@exchange.microsoft.com}

\author{Nick Craswell}
\affiliation{%
  \institution{Microsoft}
}
\email{nickcr@microsoft.com}

\newcommand{\dataset}{MIMICS\xspace}
\newcommand{\datasetone}{\dataset-Click\xspace}
\newcommand{\datasettwo}{\dataset-ClickExplore\xspace}
\newcommand{\datasetthree}{\dataset-Manual\xspace}

\begin{document}

\title{\dataset: A Large-Scale Data Collection for Search Clarification}

\begin{abstract}
Search clarification has recently attracted much attention due to its applications in search engines. It has also been recognized as a major component in conversational information seeking systems. Despite its importance, the research community still feels the lack of a large-scale data for studying different aspects of search clarification. In this paper, we introduce \dataset, a collection of search clarification datasets for real web search queries sampled from the Bing query logs. Each clarification in \dataset is generated by a Bing production algorithm and consists of a clarifying question and up to five candidate answers. \dataset contains three datasets: (1)~\datasetone includes over 400k unique queries, their associated clarification panes, and the corresponding aggregated user interaction signals (i.e., clicks). (2)~\datasettwo is an exploration data that includes aggregated user interaction signals for over 60k unique queries, each with multiple clarification panes. (3)~\datasetthree includes over 2k unique real search queries. Each query-clarification pair in this dataset has been manually labeled by at least three trained annotators. It contains graded quality labels for the clarifying question, the candidate answer set, and the landing result page for each candidate answer. 

\dataset is publicly available for research purposes,\footnote{\dataset is available at \url{https://github.com/microsoft/MIMICS}.} thus enables researchers to study a number of tasks related to search clarification, including clarification generation and selection, user engagement prediction for clarification, click models for clarification, and analyzing user interactions with search clarification.

\end{abstract}
\maketitle


\section{Introduction}
\label{sec:intro}
Search clarification has recently been recognized as a useful feature for improving user experience in search engines, especially for ambiguous and faceted queries~\cite{Zamani:2020:WWW}. In addition, it has been identified as a necessary step towards developing mixed-initiative conversational search systems~\cite{Radlinski:2017,Zamani:2020:Macaw}. The reason is that limited bandwidth interfaces used in many conversational systems, such as speech-only and small-screen devices, make it difficult or even impossible for users to go through multiple documents in case of ambiguous or faceted queries. This has recently motivated researchers and practitioners to investigate possible approaches to \emph{clarify} user information needs by asking a question~\cite{Aliannejadi:2019,Zamani:2020:WWW}.

\begin{figure}[t]
    \centering
    \includegraphics[width=\linewidth]{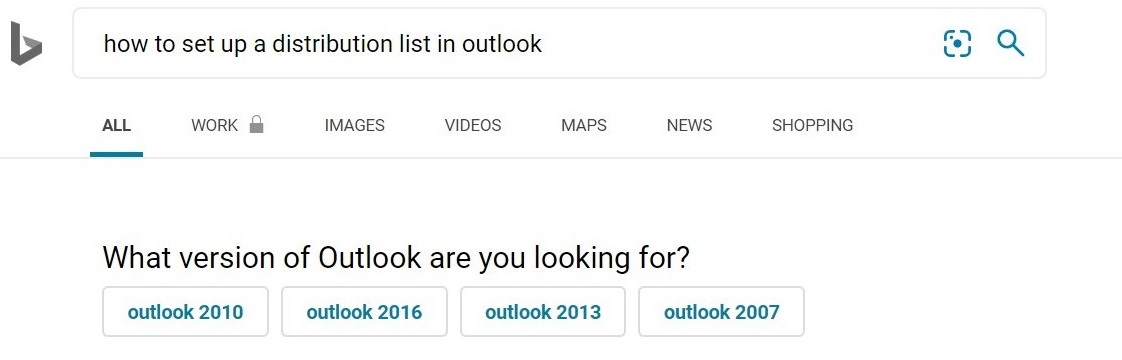}
    \caption{An example of clarification pane in Bing.}
    \label{fig:bing}
    \vspace{-0.5cm}
\end{figure}

Despite the recent progress in search clarification, e.g.,~\cite{Aliannejadi:2019,Zamani:2020:WWW,Zamani:2020:SIGIR,Hashemi:2020:SIGIR}, the community still feels the lack of a large-scale dataset for search clarification, which is necessary for speeding up the research progress in this domain.
To address this issue, we introduce \dataset,\footnote{\dataset stands for the \underline{Mi}crosoft's \underline{M}ixed-\underline{I}nitiative \underline{C}onversation \underline{S}earch Data.} a data collection consisting of multiple datasets for search clarification. Each clarification in \dataset consists of a clarifying question and up to five candidate answers. \figurename~\ref{fig:bing} shows the interface used for clarification in Bing for constructing this data. The first dataset, called \datasetone, includes over 400k unique search queries sampled from the Bing's query logs, each associated with a single clarification pane. The dataset also includes aggregated user interaction signals, such as the overall user engagement level and conditional clickthrough rate on individual candidate answers. The second dataset, called \datasettwo, contains over 64k queries, each with multiple clarification panes which are the result of multiple exploration and online randomization experiments. This dataset also includes the aggregated user interaction signals. The third dataset, on the other hand, is manually labeled by trained annotators. This dataset, which is called \datasetthree, includes graded quality labels for clarifying question, candidate answer set, and the landing result page for each individual answer.

The datasets created as part of \dataset can be used for training and evaluating a variety of tasks related to search clarification, including generating/selecting clarifying questions and candidate answers, re-ranking candidate answers for clarification, click models for search clarification, user engagement prediction for search clarification, and analyzing user interactions with search clarification. This paper also suggests some evaluation methodologies and metrics for these tasks.


\section{Related Work}
\label{sec:rel}
Clarification has been explored in a number of applications, such as speech recognition~\cite{Stoyanchev:2014}, dialogue systems~\cite{DeBoni:2003,DeBoni:2005,Quintano:2008}, and community question answering~\cite{Braslavski:2017,Rao:2018,Rao:2019}. Recently, it attracted much attention in the information retrieval literature~\cite{Aliannejadi:2019,Zamani:2020:WWW,Zamani:2020:SIGIR,Hashemi:2020:SIGIR,Radlinski:2017}. For instance, \citet{Kiesel:2018} investigated the impact of voice query clarification on user satisfaction. Their study showed that users like to be prompted for clarification. Simple form of clarification, such as entity disambiguation, has been explored by \citet{Coden:2015}. They basically ask a ``did you mean A or B?'' question to resolve entity ambiguity. Even earlier, \citet{Allan:2004} organized the HARD Track at TREC 2004 which involved clarification from participants. In more detail, the participants could submit a form containing some human-generated clarifying questions in addition to their submission run. Recently, \citet{Aliannejadi:2019} proposed studying clarification in the context of conversational information seeking systems. This was later highlighted as an important aspect of conversational search in the Dagstuhl Seminar on Conversational Search~\cite{Anand:2020}. More recently, \citet{Zamani:2020:WWW} introduced clarification in the context of web search and proposed models for generating clarifying questions and candidate answers for open-domain search queries. In a follow-up study, \citet{Zamani:2020:SIGIR} analyzed user interactions with clarification panes in Bing and provided insights into user behaviors and click bias in the context of search clarification. Moreover, \citet{Hashemi:2020:SIGIR} proposed a representation learning model for utilizing user responses to clarification in conversational search systems.

\begin{table*}[t]
    \centering
    \caption{Statistics of the datasets constructed as part of \dataset.}
    \vspace{-0.3cm}
    \renewcommand{\tabcolsep}{3pt}
    \begin{tabular}{llll}\toprule
        & \textbf{\datasetone} & \textbf{\datasettwo} & \textbf{\datasetthree} \\\midrule
        \# unique queries & 414,362 & 64,007 & 2464 \\
        \# query-clarification pairs & 414,362 & 168,921 & 2832 \\
        \# clarifications per query & 1 $\pm$ 0 & 2.64 $\pm$ 1.11 & 1.15 $\pm$ 0.36 \\
        min \& max clarifications per query & 1 \& 1 & 2 \& 89 & 1 \& 3 \\
        \# candidate answers & 2.81 $\pm$ 1.06 & 3.47 $\pm$ 1.20 & 3.06 $\pm$ 1.05 \\
        min \& max \# candidate answers & 2 \& 5 & 2 \& 5 & 2 \& 5\\
        \# query-clarification pairs with positive engagement & 71,188 & 89,441 & N/A \\
        \# query-clarification pairs with low/medium/high impressions & 264,908 / 105,879 / 43,575 & 52,071 / 60,907 / 55,943 & N/A \\
        \bottomrule
    \end{tabular}
    \label{tab:stats}
\end{table*}

Despite the recent progress reviewed above, there is no large-scale publicly available resource for search clarification. To the best of our knowledge, Qulac\footnote{\url{https://github.com/aliannejadi/qulac}}~\cite{Aliannejadi:2019} is the only public dataset that focuses on search clarification. However, it only contains 200 unique queries borrowed from the TREC Web Track 2009-2012. Therefore, it is not sufficient for training a large number of machine learning models with millions of parameters. In addition, it was constructed through crowdsourcing. Therefore, the clarifications are human generated and user responses to clarifications in real scenarios may differ from the ones in Qulac. There also exist a number of community question answering data and product catalogs with clarifications (e.g., see \cite{Rao:2019}), however, they are fundamentally different from search clarification. Therefore, this paper provides a unique resource in terms of realisticness, size, diversity, clarification types, user interaction signals, and coverage.

It is worth noting that a number of datasets related to conversational search has recently been created and released. They include CCPE-M~\cite{Radlinski:2019}, CoQA~\cite{Reddy:2019}, QuAC~\cite{Choi:2018}, MISC~\cite{Thomas:2017}, and the Conversation Assistance Track data created in TREC 2019~\cite{Dalton:2020:CAST}. Although these datasets do not particularly focus on clarification, there might be some connections between them and \dataset that can be used in future research. In addition, the public query logs, such as the one released by AOL~\cite{Pass:2006}, can be used together with \dataset for further investigations. This also holds for the datasets related to query suggestion and query auto-completion.





\section{Data Collection}
\label{sec:data}
Bing has recently added a clarification pane to its result pages for some ambiguous and faceted queries. It is located right below the search bar and above the result list. Each clarification pane includes a clarifying question and up to five candidate answers. The user interface for this feature is shown in \figurename~\ref{fig:bing}. The clarifying questions and candidate answers have been generated using a number of internal algorithms and machine learning models. They are mainly generated based on users' past interactions with the search engine (e.g., query reformulation and click), content analysis, and a taxonomy of entity types and relations. For more information on generating clarification panes, we refer the reader to~\cite{Zamani:2020:WWW} that introduces three rule-based and machine learning models for the task.  All the datasets presented in this paper follow the same properties and only demonstrate the queries from the en-US market.

In the following subsections, we explain how we created and pre-processed each dataset introduced in the paper. In summary, \dataset consists of two datasets (\datasetone and \datasettwo) based on user interactions (i.e., clicks) in Bing and one dataset (\datasetthree) based on manual annotations of clarification panes by multiple trained annotators.

\subsection{\datasetone}
\label{sec:data:one}
We sub-sampled the queries submitted to Bing in September 2019. We only kept the queries for which a clarification pane was rendered in the search engine result page (SERP). We made efforts in our data sampling to cover a diverse set of query and clarification types in the dataset, therefore, the engagement levels released in the paper by no mean represent the overall clickthrough rates in Bing. For privacy reasons, we followed $k$-anonymity by only including the queries that have been submitted by at least 40 users in the past year. In addition, the clarification panes were solely generated based on the submitted queries, therefore they do not include session and personalized information. We performed additional filtering steps to preserve the privacy of users using proprietary algorithms. Sensitive and inappropriate contents have automatically been removed from the dataset.
To reduce the noise in the click data, we removed the query-clarification pairs with less than 10 impressions. In other words, all the query-clarification pairs released in the dataset have been presented at least 10 times to the Bing users in the mentioned time period (i.e., one month). 

This resulted in 414,362 unique queries, each associated with exactly one clarification pane. Out of which 71,188 of clarifications have received positive clickthrough rates. The statistics of this dataset is presented in \tablename~\ref{tab:stats}.

The dataset is released in a tab-separated format (TSV). Each data point in \datasetone is a query-clarification pair, its impression level (low, medium, or high), its engagement level (between 0 and 10), and the conditional click probability for each individual candidate answer. The engagement level 0 means there was no click on the clarification pane. We used a equal-depth method to divide all the positive clickthrough rates into ten bins (from 1 to 10). The description of each column in the dataset is presented in \tablename~\ref{tab:clickdataformat}.

\begin{table*}[t]
    \centering
    \caption{The data format in \datasetone and \datasettwo.}
    \vspace{-0.3cm}
    \begin{tabular}{llL{11.5cm}}\toprule
        \textbf{Column(s)} & \textbf{Type} & \textbf{Description} \\\midrule
        query & string & The query text \\
        question & string & The clarifying question \\
        option\_1, $\cdots$, option\_5 & string & The candidate answers from left to right. If there is less than five candidate answers, the rest would be empty strings. \\
        impression\_level & string & A string associated with the impression level of the corresponding query-clarification pair. Its value is either 'low', 'medium', or 'high'. \\
        engagement\_level & integer  & An integer from 0 to 10 showing the level of total engagement received by the users in terms of clickthrough rate. \\
        option\_cctr\_1, $\cdots$, option\_cctr\_5 & real  & The conditional click probability on each candidate answer. They must sum to 1, unless the total\_ctr is zero. In that case, they all are zero. \\\bottomrule
    \end{tabular}
    \label{tab:clickdataformat}
\end{table*}

\subsection{\datasettwo}
\label{sec:data:two}
Although \datasetone is a invaluable resource for learning to generate clarification and related research problems, it does not allow researchers to study some tasks, such as studying click bias in user interactions with clarification. Therefore, to foster research in these interesting and practical tasks, we created \datasettwo using some exploration and randomization experiments in September 2019. In more detail, we used the top $m$ clarifications generated by our algorithms and presented them to different sets of users (similar to A/B testing). The user interactions with multiple clarification panes for the same query at the same time period enable comparison of these clarification panes. The difference between these clarification panes can be in the clarifying question, the candidate answer set, the order of candidate answers, or a combination of them.

We performed the same filtering approach to address privacy concerns as the one discussed above for \datasetone. Again, we only kept the query-clarification pairs with a minimum impression of 10. The resulted dataset contains 64,007 unique queries and 168,921 query-clarification pairs. Out of which, 89,441 query-clarification pairs received positive engagements.

The format of this dataset is the same as \datasetone (see \tablename~\ref{tab:clickdataformat}). Note that the sampling strategies for \datasetone and \datasettwo are different which resulted in significantly more query-clarification pairs with low impressions in \datasetone.

\begin{table*}[t]
    \centering
    \caption{The data format in \datasetthree. All the labels in this dataset are either 2 (Good), 1 (Fair), or 0 (Bad).}
    \vspace{-0.3cm}
    \begin{tabular}{llL{11cm}}\toprule
        \textbf{Column(s)} & \textbf{Type} & \textbf{Description} \\\midrule
        query & string & The query text \\
        question & string & The clarifying question \\
        option\_1, $\cdots$, option\_5 & string & The candidate answers from left to right. If there is less than five candidate answers, the rest would be empty strings. \\
        question\_label & integer & The label associated with the clarifying question independent of the candidate answers. \\
        options\_overall\_label & integer & The overall label given to the candidate answer set. \\
        option\_label\_1, $\cdots$, option\_label\_5 & integer  & The label assigned to each individual candidate answer based on the quality of the landing search result page. \\\bottomrule
    \end{tabular}
    \label{tab:labeldataformat}
\end{table*}




\begin{table*}[t]
    \centering
    \caption{The statistics of the common clarifying question templates in \dataset. We only present the templates with at least 100 occurrence in \datasetone and \datasettwo individually. Note that there is no label associated with the first template in \datasetthree.}
    \vspace{-0.3cm}
    \renewcommand{\tabcolsep}{2pt}
    \begin{tabular}{llllllll}\toprule
        \multirow{2}{*}{\textbf{ID}} & \multirow{2}{*}{\textbf{Clarifying question template}} & \multicolumn{2}{c}{\textbf{\datasetone}} & \multicolumn{2}{c}{\textbf{\datasettwo}} & \multicolumn{2}{c}{\textbf{\datasetthree}} \\
        & & \textbf{Freq.} & \textbf{Engagement}  & \textbf{Freq.} & \textbf{Engagement}  & \textbf{Freq.} & \textbf{Question Quality} \\\midrule
        T1 & select one to refine your search &	395134 &	0.9285 &	156870 &	2.8631 &  2490 & N/A \\
        T2 & what (do you want|would you like) to know about (.+)? &	7136 &	0.5783 &	5624 &	2.9070 & 158 & 1.9367 \\
        T3 & (which|what) (.+) do you mean? &	7483 &	0.6123 &	1905 &	2.6714 & 76 & 2.000 \\
        T4 & (what|which) (.+) are you looking for? 	& 3436 &	1.7252	& 2055 &	5.1990 & 22 & 1.6818 \\
        T5 & what (do you want|would you like) to do with (.+)? &	689 &	1.9637 &	1833 &	3.4043 & 60 & 2.000 \\
        T6 & who are you shopping for? 	& 101 &	1.9604 &	350	 & 4.3800 & 7 & 1.5714 \\
        T7 & what are you trying to do? & 	188 &	3.3777 &	116 &	5.8793 & 3 & 1.0\\
        \bottomrule
    \end{tabular}
    \label{tab:templates}
\end{table*}

\begin{table}[t]
    \centering
    \caption{The average and standard deviation of user engagement levels with respect to different query-clarification impressions.}
    \vspace{-0.3cm}
    \renewcommand{\tabcolsep}{4pt}
    \begin{tabular}{lll}\toprule
        \textbf{Impression level} & \textbf{\datasetone} & \textbf{\datasettwo} \\\midrule
        Low & 0.9061 $\pm$ 2.5227 & 3.1712 $\pm$ 4.2735 \\
        Medium & 0.9746 $\pm$ 2.1249 & 3.1247 $\pm$ 3.3622 \\
        High & 0.9356 $\pm$ 1.6159 & 2.4119 $\pm$ 2.4559 \\\bottomrule
    \end{tabular}
    \label{tab:impression}
\end{table}

\subsection{\datasetthree}
\label{sec:data:three}
Although click provides a strong implicit feedback signal for estimating the quality of models in online services, including search clarification, it does not necessarily reflect all quality aspects. In addition, it can be biased for many reasons. Therefore, a comprehensive study of clarification must include evaluation based on manual human annotations. This has motivated us to create and release \datasetthree based on manual judgements performed by trained annotators. 

Therefore, we randomly sampled queries from the query logs to collect manual annotations for a set of realistic user queries. The queries satisfy all the privacy concerns reviewed in Section~\ref{sec:data:one}. We further used the same algorithm to generate one or more clarification pairs for each query. Each query-clarification pair was assigned to at least three annotators. The annotators have been trained to judge clarification panes by attending online meetings, reading comprehensive guidelines, and practicing. In the following, we describe each step in the designed Human Intelligence Task (HIT) for annotating a query-clarification pair. This guideline has been previously used in \cite{Zamani:2020:WWW,Zamani:2020:SIGIR}.

\subsubsection{Step I: SERP Review}
Similar to \citet{Aliannejadi:2019}, we first asked the annotators to skim and review a few pages of the search results returned by Bing. Since search engines try to diversify the result lists, this would enable the annotators to better understand the scope of the topic and different potential intents behind the submitted query. When completed, the users can move to the next step.

\subsubsection{Step II: Annotating the Clarifying Question Quality}
In this step, the annotators were asked to assess the quality of the given clarifying question independent of the candidate answers. Therefore, the annotation interface does not show the candidate answers to the annotators at this stage. Each clarifying question is given a label 2 (Good), 1 (Fair), or 0 (Bad). The annotators were given detailed definitions, guidelines, and examples for each of the labels. In summary, the guideline indicates that a Good clarifying question should accurately address and clarify different intents of the query. It should be fluent and grammatically correct. If a question fails in satisfying any of these factors but still is an acceptable clarifying question, it should be given a Fair label. Otherwise, a Bad label should be assigned to the question. Note that if a question contains sensitive or inappropriate content, it would have been flagged by the annotators and removed from the dataset. Note that in case of having a generic template instead of clarifying questions (i.e., ``select one to refine your search''), we do not ask the annotators to provide a question quality labels.

\subsubsection{Step III: Annotating the Candidate Answer Set Quality}
Once the clarifying question is annotated, the candidate answers would appear on the HIT interface. In this step, the annotators were asked to judge the overall quality of the candidate answer set. In summary, the annotation guideline indicates that the candidate answer set should be evaluated based on its usefulness for clarification, comprehensiveness, coverage, understandability, grammar, diversity, and importance order. A clear definition of each of these constraints has been mentioned in the guideline. 
Note that the annotators have reviewed multiple pages of the result list in Step I and have been expected to know different possible intents of the query. Again, the labels are either 2 (Good), 1 (Fair), or 0 (Bad), and the candidate answers with sensitive or inappropriate contents have been removed from the dataset. If a candidate answer set satisfies all the aforementioned constraints, it should be given a Good label. While, the Fair label should be given to an acceptable  candidate answer set that does not satisfy at least one of the constraints. Otherwise, the Bad label should be chosen. Note that since all the defined properties are difficult to satisfy with up to 5 candidate answers, the label Good is rarely chosen for a candidate answer set.





\begin{table*}[t]
    \centering
    \caption{The average and standard deviation of engagement levels and manual annotation labels per query length.}
    \vspace{-0.3cm}
    \renewcommand{\tabcolsep}{3pt}
    \begin{tabular}{lllllllll}\toprule
        \textbf{Query} & \multicolumn{2}{c}{\textbf{\datasetone}} & \multicolumn{2}{c}{\textbf{\datasettwo}} & \multicolumn{4}{c}{\textbf{\datasetthree}} \\
        \textbf{length} & \textbf{Freq.} & \textbf{Engagement} & \textbf{Freq.} & \textbf{Engagement} & \textbf{Freq.} & \textbf{Question quality} & \textbf{Answer set quality} & \textbf{Landing page quality} \\\midrule
        1 &	52213 &	0.5158 $\pm$ 1.6546 &	26926 &	1.9508 $\pm$ 2.7098 & 1028 & 1.7347 $\pm$ 0.4415 & 1.0418 $\pm$ 0.3075 & 1.9750 $\pm$ 0.1251\\
        2 &	160161 &	0.7926 $\pm$ 2.1548 &	70621 &	2.7965 $\pm$ 3.3536 & 942 & 1.4694 $\pm$ 0.4991 & 1.0085 $\pm$ 0.3827 & 1.9178 $\pm$ 0.2881\\
        3 &	120821 & 1.0152 $\pm$ 2.4573 &	46070 &	3.1677 $\pm$ 3.5811 & 555 & 1.4667 $\pm$ 0.4989 & 0.9333 $\pm$ 0.4463 & 1.8021 $\pm$ 0.4816\\
        4 &	51503 &	1.2196 $\pm$ 2.6980 &	16798 &	3.5397 $\pm$ 3.7492 & 199 & 1.3333 $\pm$ 0.4714 & 0.9698 $\pm$ 0.5103 & 1.8313 $\pm$ 0.41986\\
        5 &	19893 &	1.4473 $\pm$ 2.9078	& 5755 &	4.0188 $\pm$ 3.8921 & 75 & 1.3846 $\pm$ 0.4865 & 1.0267 $\pm$ 0.5157 & 1.7847 $\pm$ 0.5291\\
        6 &	6299 &	1.5785 $\pm$ 3.0318 &	1806 &	4.1877 $\pm$ 3.9642 & 15 & 1.0 $\pm$ 0.0 & 0.8 $\pm$ 0.5416 & 1.7 $\pm$ 0.4800\\
        7 &	2424 &	1.6634 $\pm$ 3.0815	& 621 &	4.6715 $\pm$ 3.9861 & 13 & 1.0 $\pm$ 0.0 & 0.7692 $\pm$ 0.4213 & 1.7692 $\pm$ 0.5756\\
        8 &	823 & 1.7618 $\pm$ 3.1575 &	264 & 4.2008 $\pm$ 3.9082 & 3 & N/A & 1.0 $\pm$ 0.0 & 1.8333 $\pm$ 0.2357\\
        9 &	184 & 1.9620 $\pm$ 3.2959 &	52 & 4.1731 $\pm$ 3.8467 & 1 & N/A & 0.0 $\pm$ 0.0 & 2.0 $\pm$ 0.0\\
        10+ &	41 & 2.0732 $\pm$ 3.4244 & 8 & 4.8750 $\pm$ 3.4799 & 1 & N/A & 1.0 $\pm$ 0.0 & 2.0 $\pm$ 0.0\\
        \bottomrule
    \end{tabular}
    \label{tab:qlen}
\end{table*}

\subsubsection{Step IV: Annotating the Landing SERP Quality for Each Individual Candidate Answer}
\citet{Zamani:2020:WWW} recently performed a number of user studies related to search clarification. In their interviews, the participants mentioned that the quality of the secondary result page (after clicking on a candidate answer) perceived the usefulness of the clarification pane. Based on this observation, we asked the annotators to evaluate the quality of the secondary result page (or the landing result page) for the individual candidate answers one by one. Therefore, the annotators could click on each individual answer and observe the secondary result page in Bing. Since a SERP may contain multiple direct answers, entity cards, query suggestion, etc. in addition to the list of webpages, adopting ranking metrics based on document relevance, such as mean reciprocal rank (MRR) or normalized discounted cumulative gain (NDCG)~\cite{Jarvelin:2002}, is not desired to evaluate the overall SERP quality. Therefore, we again asked the annotators to assign a label 2 (Good), 1 (Fair), or 0 (Bad) to each landing SERP. A label Good should be chosen, if the correct answer to all possible information needs behind the selected candidate answer can be easily found in a prominent location in the page (e.g., an answer box on top of the SERP or the top three retrieved webpages). If the result page is still useful and contain relevant information, but finding the answer is not easy or is not on top of the SERP, the Fair label should be selected. Otherwise, the landing SERP should be considered as Bad.

\subsubsection{A Summary of the Collected Data}
Each HIT was assigned to at least three annotators. For each labeling task, we used majority voting to aggregate the annotation. In case of disagreements, the HIT was assigned to more annotators. The overall Fleiss’ kappa inter-annotator agreement is 63.23\%, which is considered as good. 

Our annotations resulted in over 2.4k unique queries and over 2.8k query-clarification pairs. The statistics of the dataset is reported in \tablename~\ref{tab:stats}.
The data has been released in a tab-separated file format (TSV). The description of each column in the data is provided in \tablename~\ref{tab:labeldataformat}.

\section{Data Analysis}
\label{sec:analysis}
In this section, we provide a comprehensive analysis of the created datasets.

\subsection{Question Template Analysis}
\citet{Zamani:2020:WWW} showed that most search clarifications can be resolved using a small number of question templates. In our first set of analysis, we study the question templates in \dataset and their corresponding statistics. We only focus on the templates with a minimum frequency of 100 in both \datasetone and \datasettwo. We compute the average engagement level per clarifying question template for \datasetone and \datasettwo. In addition, we compute the average question quality label per template for \datasetthree that has manual annotations. Note that engagement levels are in the $[0, 10]$ interval, while the manual annotation labels are in $[0, 2]$. The results are reported in \tablename~\ref{tab:templates}. The first general template is excluded in our manual annotations. According to the results, the last four templates (T4 - T7) have led to higher engagements compared to T1, T2, and T3 in both \datasetone and \datasettwo. They are also generally less frequent in the dataset and more specific. In general, the exploration dataset has higher average engagements compared to \datasetone. The reason is that the number of query-clarification pairs with zero engagements in \datasetone are higher than those in \datasettwo (see \tablename~\ref{tab:stats}).

\subsection{Analyzing Engagement Based on Clarification Impression}
As mentioned in Section~\ref{sec:data}, \datasetone and \datasettwo contain a three-level impression label per query-clarification pair. The impression level is computed based on the number of times the given query-clarification pair has been presented to users. The impression level should have a correlation with the query frequency. We compute the average and standard deviation of engagements per impression level whose results are reported in \tablename~\ref{tab:impression}. According to the results, there is a negligible difference between the average engagements across impression levels. Given the engagements range (i.e., $[0, 10]$), the query-clarification pairs with high impressions in \datasettwo have led to slightly lower average engagements.

\subsection{Analysis Based on Query Length}
In our third analysis, we study user engagements and manual quality labels with respect to query length. To this aim, we compute the query length by simply splitting the query using whitespace characters as delimiters. The results are reported in \tablename~\ref{tab:qlen}. According to the results on \datasetone and \datasettwo, the average engagement increases as the queries get longer. By looking at the data one can realize that longer queries are often natural language questions, while short queries are keyword queries. Surprisingly, this is inconsistent with the manual annotations suggesting that single word queries have higher question quality, answer set quality, and also landing page quality (excluding the rare queries with less than 10 frequency in the dataset). This observation suggests that user engagement with clarification is not necessarily aligned with the clarification quality. The behavior of users who submit longer queries may differ from those who search with keyword queries.



\begin{table*}[t]
    \centering
    \caption{The average and standard deviation of engagement levels and manual annotation labels per number of candidate answers.}
    \vspace{-0.3cm}
    \renewcommand{\tabcolsep}{2pt}
    \begin{tabular}{lllllllll}\toprule
        \multirow{2}{*}{\textbf{\# answers}} & \multicolumn{2}{c}{\textbf{\datasetone}} & \multicolumn{2}{c}{\textbf{\datasettwo}} & \multicolumn{4}{c}{\textbf{\datasetthree}} \\
        & \textbf{Freq.} & \textbf{Engagement} & \textbf{Freq.} & \textbf{Engagement} & \textbf{Freq.} & \textbf{Question quality} & \textbf{Answer set quality} & \textbf{Landing page quality} \\\midrule
        2 & 226697 & 0.9047 $\pm$ 2.3160 & 50474 & 2.8430 $\pm$ 3.3921 & 1083 & 1.3164 $\pm$ 0.4651 & 0.9751 $\pm$ 0.3775 & 1.8915 $\pm$ 0.3665\\
        3 & 91840 & 0.9904 $\pm$ 2.4175 & 38619 & 3.0592 $\pm$ 3.5111 & 892 & 1.7513 $\pm$ 0.4323 & 0.9507 $\pm$ 0.2954 & 1.9129 $\pm$ 0.3101\\
        4 & 42752 & 0.9276 $\pm$ 2.3505 & 29678 & 2.9157 $\pm$ 3.4395 & 453 & 1.6292 $\pm$ 0.4830 & 1.0088 $\pm$ 0.3816 & 1.9073 $\pm$ 0.2862\\
        5 & 53073 & 0.9099 $\pm$ 2.3323 & 50150 & 2.8354 $\pm$ 3.4236 & 404 & 1.4741 $\pm$ 0.4993 & 1.1733 $\pm$ 0.5401 & 1.9168 $\pm$ 0.2832\\
        \bottomrule
    \end{tabular}
    \label{tab:num_answerset}
\end{table*}

\subsection{Analysis Based on the Number of Candidate Answers}
As pointed out earlier, the number of candidate answers in the data varies between two and five. To demonstrate the impact of the number of candidate answers, we report the average and standard deviation of engagement levels and manual quality labels per number of candidate answers in \tablename~\ref{tab:num_answerset}. According to the results, there is a small difference between average engagements in both \datasetone and \datasettwo datasets. The clarifications with three candidate answers have led to a slightly higher engagement than the rest. It is again in contrary to the manual quality labels; the clarifications with three candidate answers have obtained the lowest answer set quality label. On the other hand, the question quality of clarifications with three candidate answers is higher than the others. This highlights that the question quality may play a key role in increasing user engagements.

\begin{figure*}
\begin{subfigure}{.25\textwidth}
  \centering
  \includegraphics[width=\linewidth]{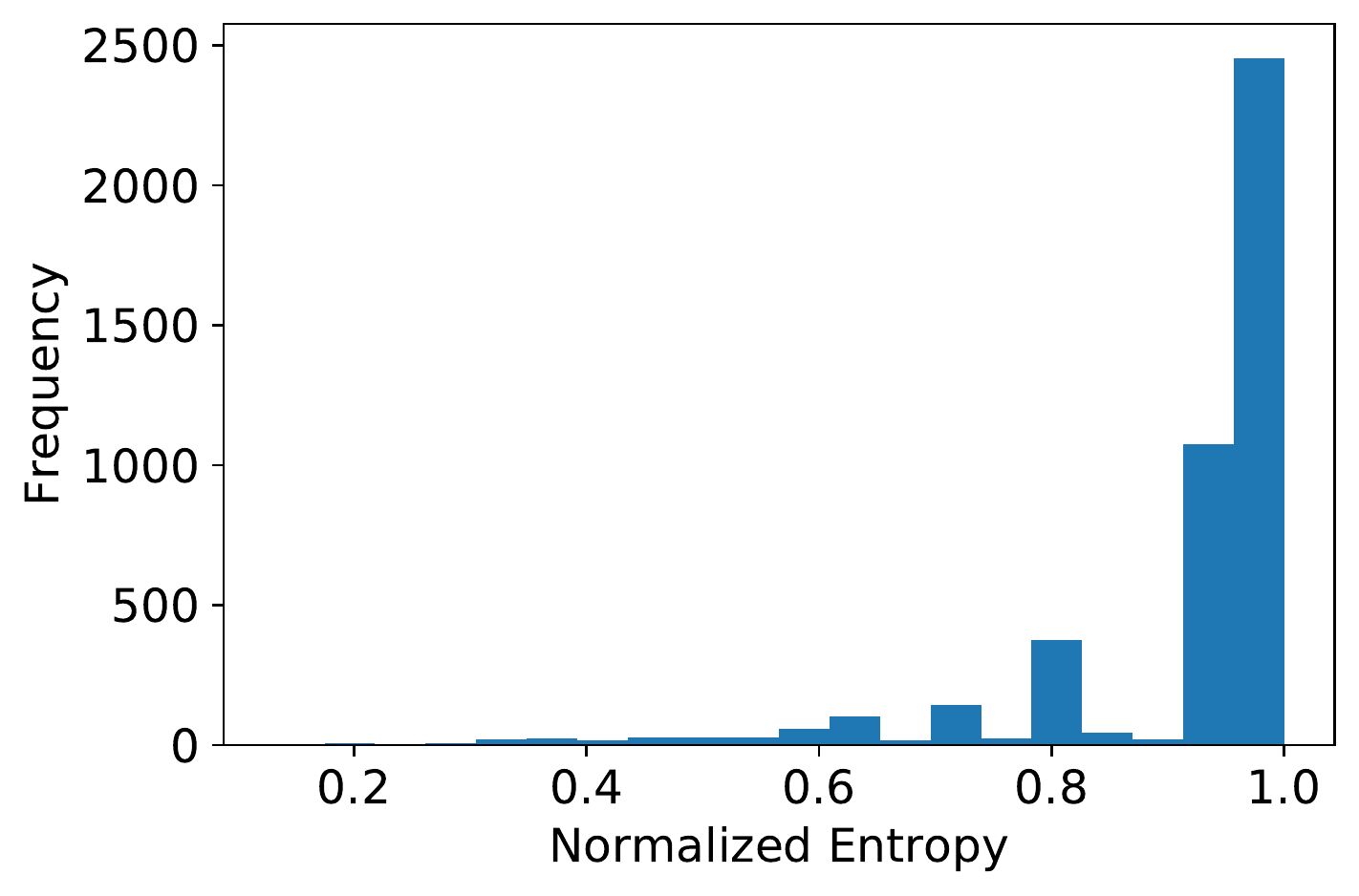}
  \caption{\# candidate answers = 2}
  \label{fig:sfig1}
\end{subfigure}%
\begin{subfigure}{.25\textwidth}
  \centering
  \includegraphics[width=\linewidth]{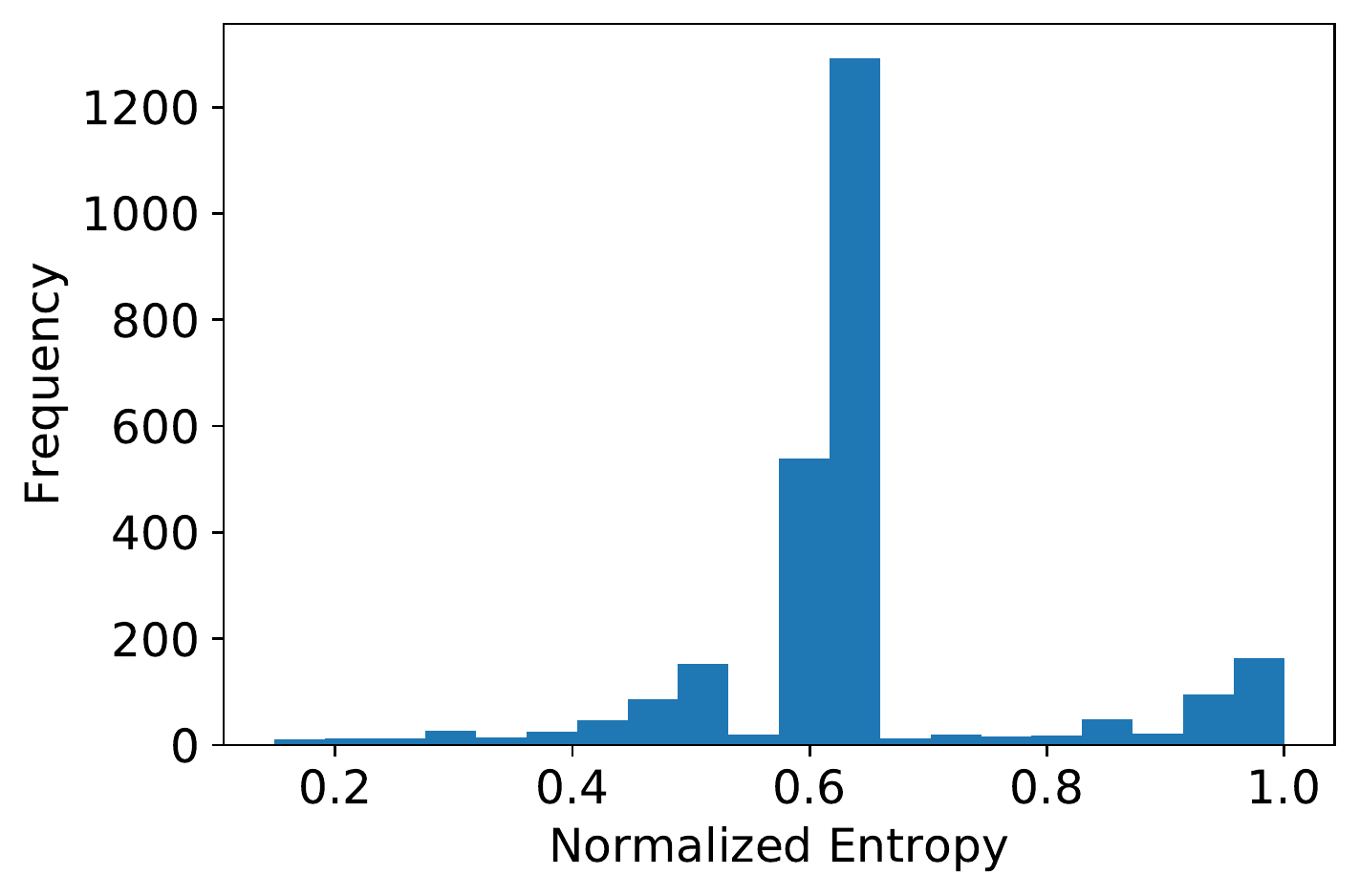}
  \caption{\# candidate answers = 3}
  \label{fig:sfig1}
\end{subfigure}%
\begin{subfigure}{.25\textwidth}
  \centering
  \includegraphics[width=\linewidth]{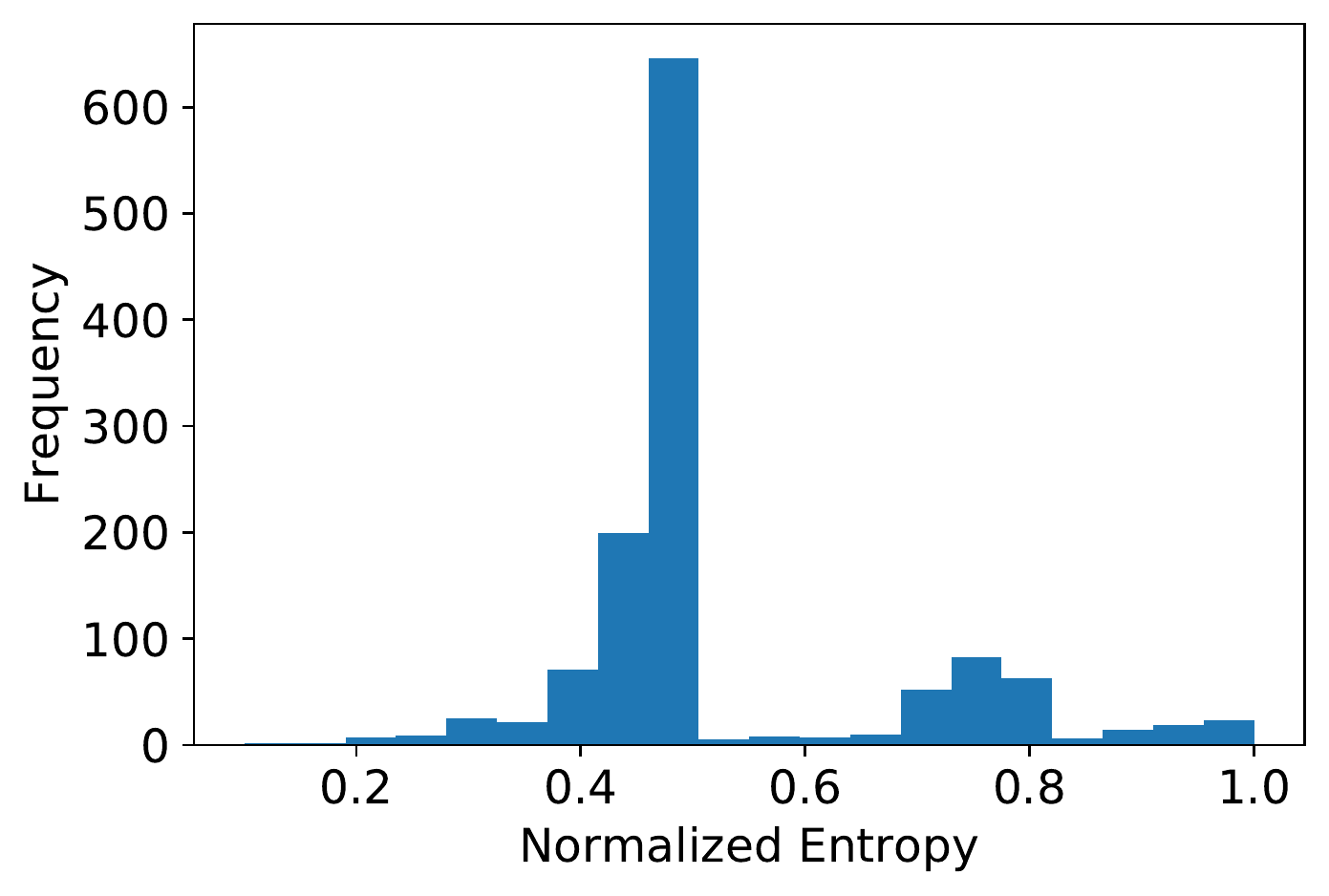}
  \caption{\# candidate answers = 4}
  \label{fig:sfig2}
\end{subfigure}%
\begin{subfigure}{.25\textwidth}
  \centering
  \includegraphics[width=\linewidth]{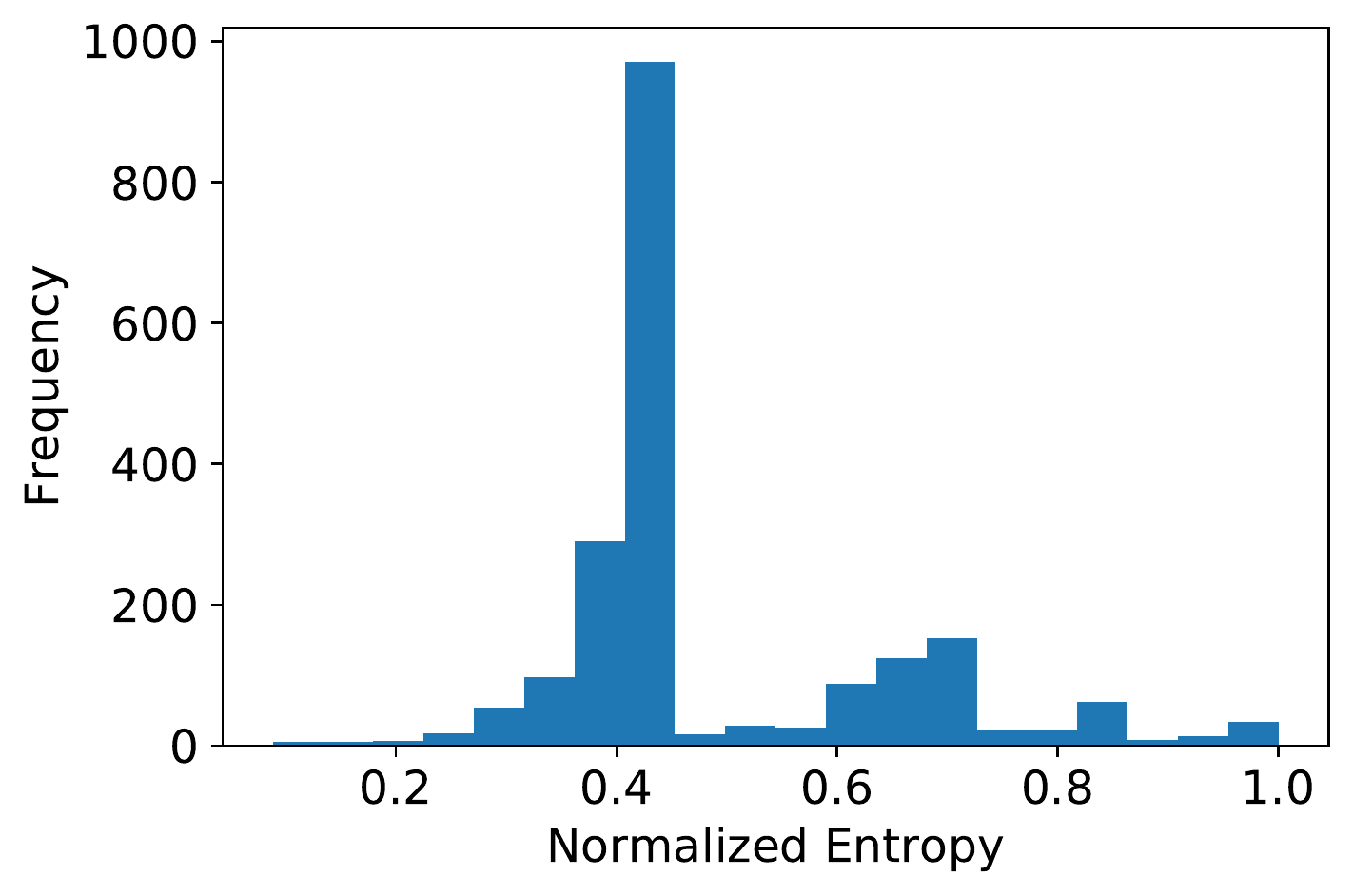}
  \caption{\# candidate answers = 5}
  \label{fig:sfig1}
\end{subfigure}%
%
\caption{The distribution of normalized entropy for the conditional clickthrough rates on candidate answers for the \datasetone dataset. For the sake of clarity and visualization, we exclude the clarification with no click and those with zero entropy.}
\label{fig:entropy:one}
\end{figure*}

\begin{figure*}
\begin{subfigure}{.25\textwidth}
  \centering
  \includegraphics[width=\linewidth]{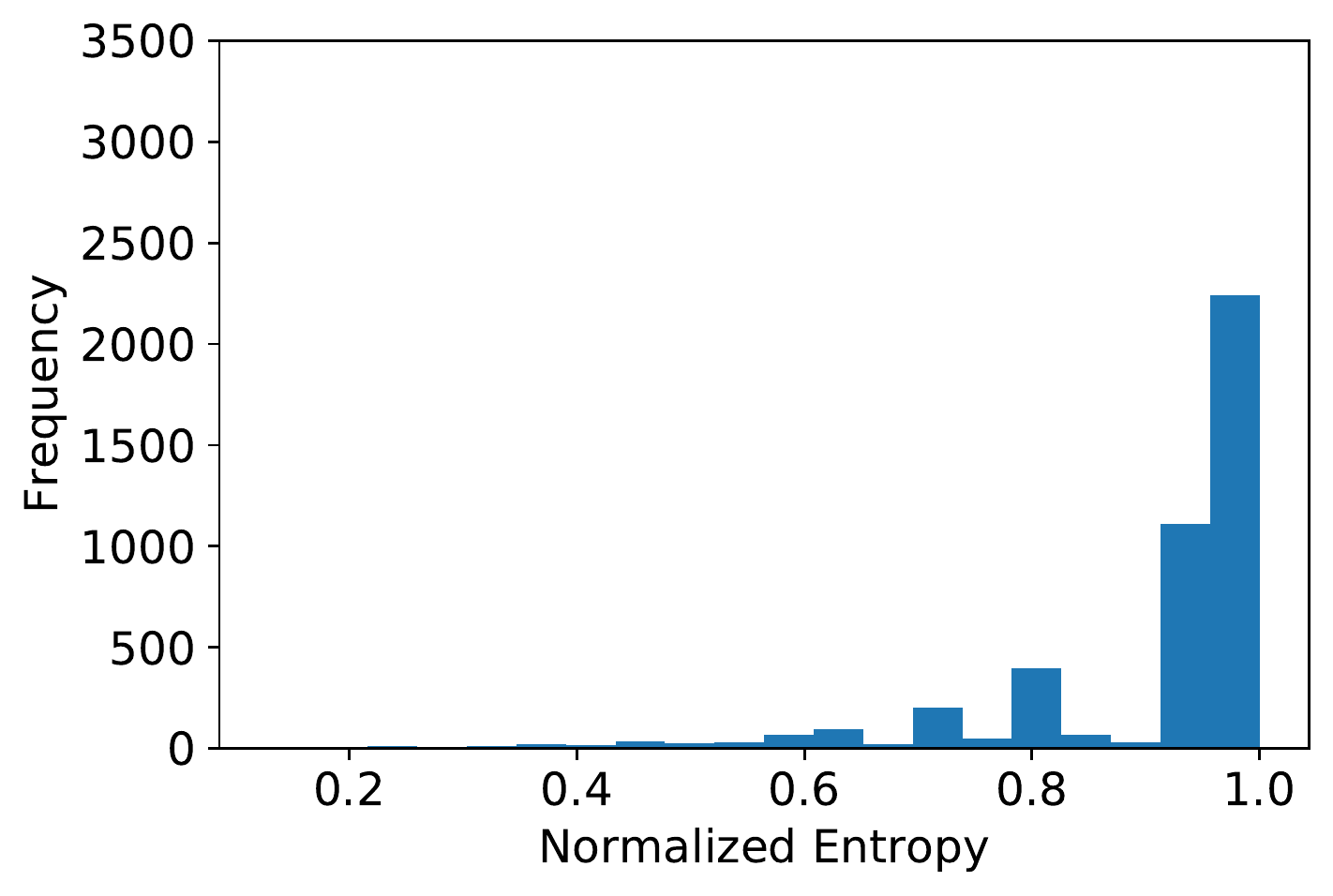}
  \caption{\# candidate answers = 2}
  \label{fig:sfig1}
\end{subfigure}%
\begin{subfigure}{.25\textwidth}
  \centering
  \includegraphics[width=\linewidth]{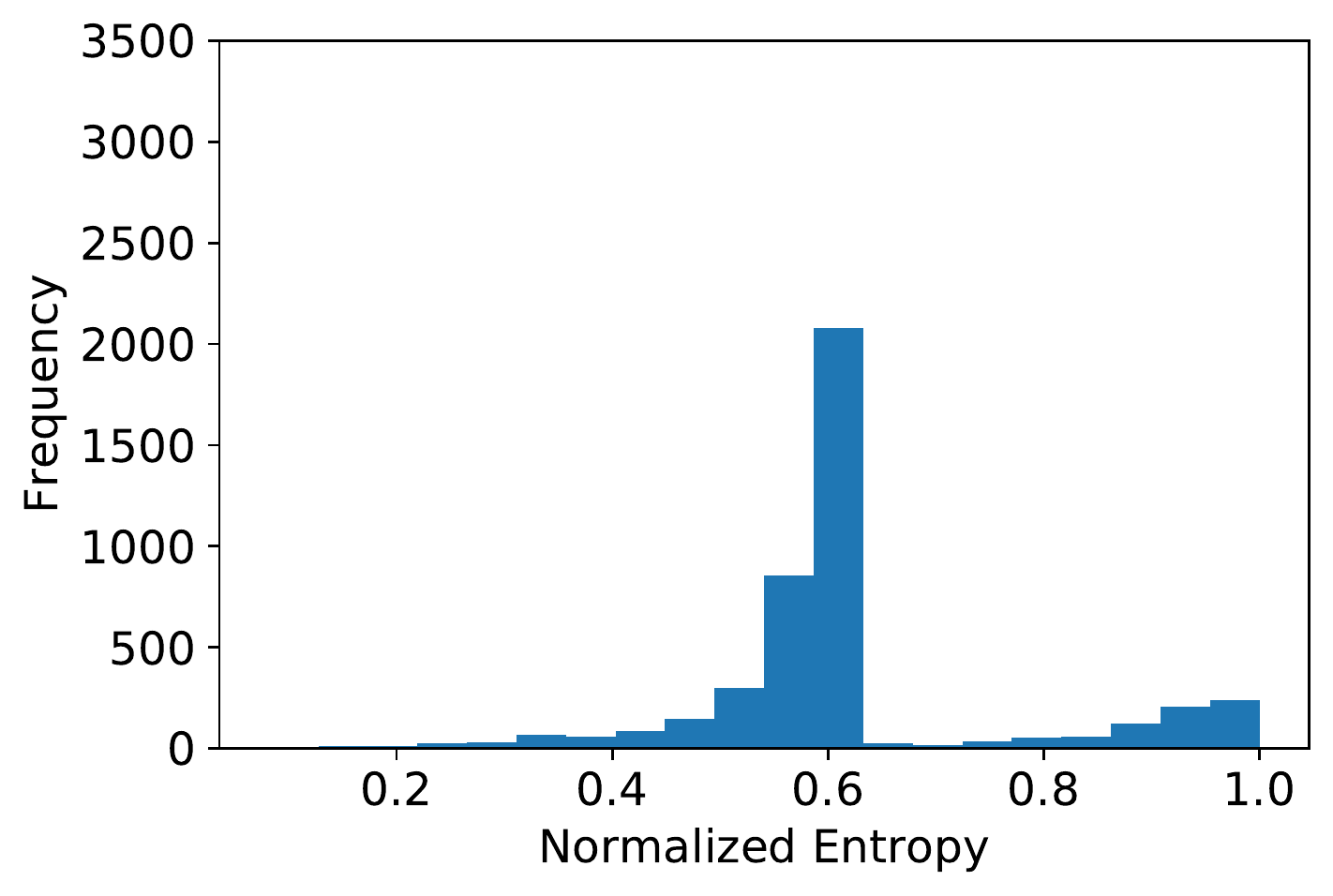}
  \caption{\# candidate answers = 3}
  \label{fig:sfig1}
\end{subfigure}%
\begin{subfigure}{.25\textwidth}
  \centering
  \includegraphics[width=\linewidth]{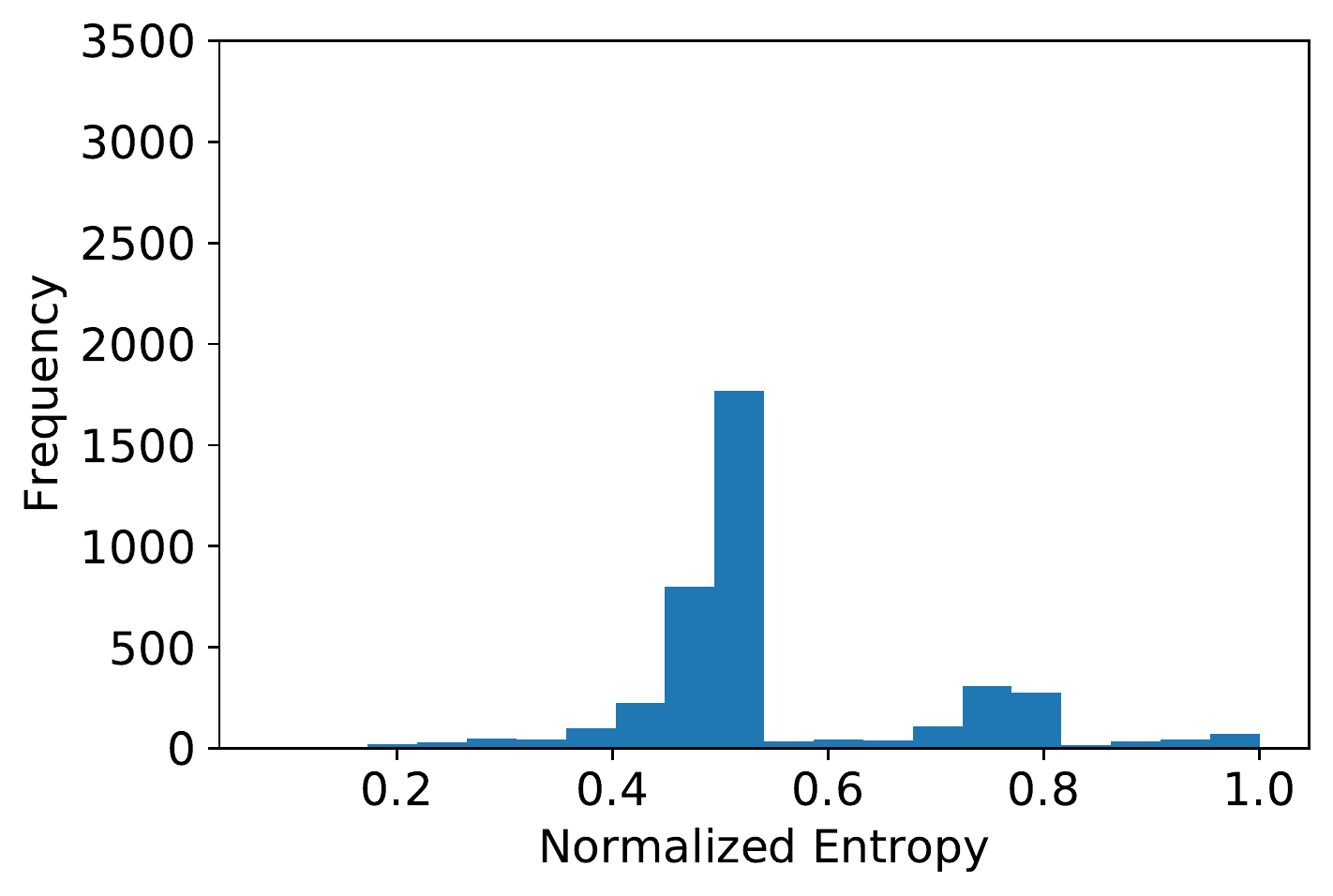}
  \caption{\# candidate answers = 4}
  \label{fig:sfig2}
\end{subfigure}%
\begin{subfigure}{.25\textwidth}
  \centering
  \includegraphics[width=\linewidth]{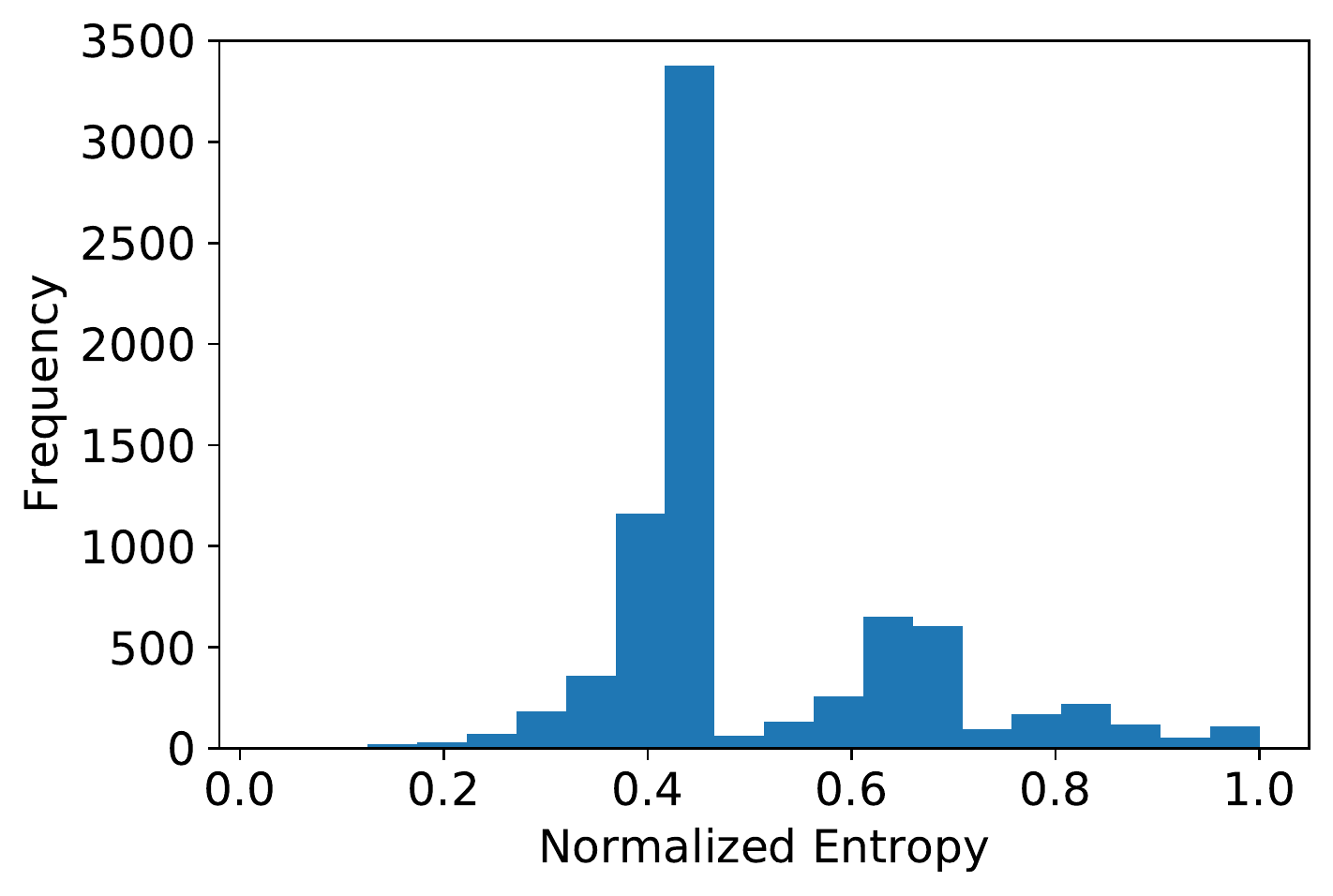}
  \caption{\# candidate answers = 5}
  \label{fig:sfig1}
\end{subfigure}%
%
\caption{The distribution of normalized entropy for the conditional clickthrough rates on candidate answers for the \datasettwo dataset. For the sake of clarity and visualization, we exclude the clarification with no click and those with zero entropy.}
\label{fig:entropy:two}
\end{figure*}

\subsection{Analyzing Click Entropy Distribution on Candidate Answers}
\datasetone and \datasettwo both contain conditional click probability on each individual answer, i.e., the probability of clicking on each candidate answer assuming that the user interacts with the clarification pane. The entropy of this probabilistic distribution demonstrates how clicks are distributed across candidate answers. The entropy range depends on the number of candidate answers, therefore, we normalized the entropy values by the maximum entropy per the candidate answer size. The distribution for \datasetone and \datasettwo are reported in Figures~\ref{fig:entropy:one} and \ref{fig:entropy:two}, respectively. Note that for the sake of visualization, these plots do not include clarifications with no click (i.e., the engagement level zero) and those with zero entropy. According to the plots, the number of peaks in the entropy distribution is aligned with the number of candidate answers. The entropy values where the histogram peaks suggest that in many cases there is a uniform-like distribution for $m$ out of $n$ candidate answers (for all values of $m$). Comparing the plots in \figurename~\ref{fig:entropy:one} with those in \figurename~\ref{fig:entropy:two} shoes that this finding is consistent across datasets.


\section{Introducing Research Problems Related to Search Clarification}
\dataset enables researchers to study a number of research problems. In this section, we introduce these tasks and provide high-level suggestions for evaluating the tasks using \dataset.

\subsection{Clarification Generation}
Clarification generation (including both clarifying question and candidate answers) is a core task in search clarification. Generating clarification from a passage-level text has been studied in the context of community question answering posts~\cite{Rao:2019}. It has lately attracted much attention in information seeking systems, such as search engines (similar to this study)~\cite{Zamani:2020:WWW} and recommender systems~\cite{Zhang:2018}. Previous work has pointed out the lack of large-scale training data for generating search clarification~\cite{Zamani:2020:WWW,Aliannejadi:2019}. \dataset, especially the click data, provides an excellent resource for training clarification generation models. 

Evaluating clarification generation models, on the other hand, is difficult. One can use \dataset for evaluating the generated clarification models using metrics such as BLEU~\cite{Papineni:2002} and ROUGE~\cite{Lin:2003}. However, we strongly discourage this evaluation methodologies, as they poorly correlate with user satisfaction and clarification quality. Here is our recommendation for evaluating clarification generation models:
\begin{itemize}[leftmargin=*]
    \item In case of access to production systems with real users, conducting online experiments (e.g., A/B tests) would be a reliable evaluation methodology and the models can be compared using user engagement measures, such as clickthrough rate.
    
    \item Manual annotation of the generated clarifications based on carefully-defined criteria would be an alternative for clarification generation evaluation. Previously, \citet{Zamani:2020:WWW} used this evaluation methodologies. Researchers may adopt the annotation guideline presented in Section~\ref{sec:data:three} for designing their crowdsourcing HITs.
\end{itemize}

\subsection{Clarification Selection}
Since automatic offline evaluation of clarification generation models is difficult, clarification selection (or clarification re-ranking) can be considered as an auxiliary task to evaluate the quality of learned representations for clarification. In addition, as pointed out by \citet{Aliannejadi:2019}, information seeking systems can adopt a two stage process for asking clarification, i.e., generating multiple clarifications and selecting one. Selecting clarification has been previously studied in \cite{Zamani:2020:SIGIR,Aliannejadi:2019,Hashemi:2020:SIGIR}.

Researchers can benefit from \dataset for both training and evaluating clarification selection models. In more detail, \datasettwo contains multiple clarifications per query and can be directly used for evaluating clarification selection (or re-ranking) models. The other two datasets can be also used by drawing some negative samples that can be obtained either randomly or using a baseline model.

Ranking metrics, such as NDCG, can be used to evaluate clarification selection models. In addition, since only one clarification is often shown to the users, the average engagement of the selected clarification can be also chosen as an evaluation metric. Refer to \cite{Zamani:2020:SIGIR} for more information.


\subsection{User Engagement Prediction for Clarification}
A major task in search clarification is deciding whether to ask clarification, especially in search systems with limited-bandwidth interfaces. This problem can be cast to query performance prediction~\cite{Cronen-Townsend:2002,Carmel:2010}. In other words, clarification can be asked when the predicted performance for the given query is below a threshold. An alternative to query performance prediction for this task would be user engagement prediction. In more detail, if users enjoy interacting with clarification and find it useful, the system can decide to ask the clarification. Predicting user engagement has been previously studied in various contexts, such as social media and web applications~\cite{Lalmas:2014,Zamani:2015}, however, user engagement prediction for clarification is fundamentally different. \datasetone and \datasettwo contain engagement levels in the $[0, 10]$ interval. Therefore, they can be directly used for predicting user engagements. 

For evaluating user engagements prediction models for clarification, we recommend computing correlation between the predicted engagements and the actual observed engagement released in the datasets. Correlation has been also used for evaluating query performance prediction models~\cite{Carmel:2010}. Since we only release engagement levels, we suggest using both linear (e.g., Pearson's $\rho$) and rank-based (e.g., Kendall's $\tau$) correlation metrics.

In addition, mean square error or mean absolute error can be used for evaluating user engagement prediction methods.

\subsection{Re-ranking Candidate Answers}
Previous work has shown that the order of candidate answers in clarification matters~\cite{Zamani:2020:SIGIR}. \dataset enables researchers to study the task of re-ranking candidate answers for a given pair of query and clarifying question. Experiments on both click data (\datasetone and \datasettwo) and manual annotations would provide complementary evaluation for the task. 

For evaluating the candidate answers re-ranking task, the manual annotations per individual answers based on their landing SERP quality can be used as graded relevance judgement. NDCG would be adopted as the evaluation metric. For evaluation using the click data, researchers should be careful about presentation bias in the data. Refer to \cite{Zamani:2020:SIGIR} for more detail. In summary, the candidate answers with higher ranks and longer text are more likely to attract clicks. This point should be considered prior to using the \datasetone and \datasettwo for re-ranking candidate answers. Once this issue is addressed, the conditional click probabilities can be mapped to ordinal relevance labels and typical ranking metrics can be adopted for evaluation. One can also use cross-entropy between the predicted probability distribution for candidate answers and the actual conditional click distribution. The impression level can be also considered in the metric to compute a gain per query-clarification pair with respect to their impression. In more detail, the clarifications that are presented more often should be assigned higher weights.

\subsection{Click Models for Clarification}
Related to the re-ranking candidate answers task, it is important to design user models for their click behavior while interacting with clarification panes. \citet{Zamani:2020:SIGIR} showed that the existing click models that have primarily been designed for web search do not perform as expected for search clarification. The reason is that the assumptions made in the web search click models do not hold for search clarification. The \datasettwo dataset contains many clarification pairs for a given query whose only differences are in the order of candidate answers. This allows researchers to train and evaluate click models for search clarification using \datasettwo. The evaluation methodology used in \cite{Zamani:2020:SIGIR} is suggested for evaluating the task. In summary, it is based on predicting the click probability of swapping adjacent candidate answers. This approach has originally been used for evaluating click models in web search by \citet{Craswell:2008}. The cross-entropy would be an appropriate metric in this evaluation setup.

\subsection{Analyzing User Behavior in Search Clarification}
Although this paper provides several analyses based on search clarification quality in terms of both manual judgements and engagement levels, future work can benefit from \datasetone and \datasettwo to conduct more in depth analysis of user behaviors while interacting with search clarification in the context of web search.

\section{Conclusions}
\label{sec:conclusions}
In this paper, we introduced \dataset, a data collection for studying search clarification, which is an interesting and emerging task in the context of web search and conversational search. \dataset was constructed based on the queries and interactions of real users, collected from the search logs of a major commercial web search engine. \dataset consists of three datasets: (1) \datasetone includes over 400k unique queries with the associated clarification panes. (2) \datasettwo is an exploration data and contains multiple clarification panes per query. It includes over 60k unique queries. (3) \datasetthree is a smaller dataset with manual annotations for clarifying questions, candidate answer sets, and the landing result page after clicking on individual candidate answers. We publicly released these datasets for research purposes.

We also conducted a comprehensive analysis of the user interactions and manual annotations in our datasets and shed light on different aspects of search clarification. We finally introduced a number of key research problems for which researchers can benefit from \dataset.

In the future, we intend to report benchmark results for a number of standard baselines for each individual task introduced in the paper. We will release the results to improve reproducibility and comparison.
There exist a number of limitations in the released datasets. For instance, they only focus on the en-US market and do not contain personalized and session-level information. These limitations can be resolved in the future.



\end{document}